\pdfoutput=1
\documentclass[12pt]{article}
\usepackage{putex}
\usepackage{graphicx}

\begin{document}


\thispagestyle{empty}

\begin{flushright}\scshape
February 2014
\end{flushright}
\vskip1cm

\begin{center}

{\LARGE\scshape The lowest resonance in QCD from low--energy data
\par}
\vskip15mm

\textsc{L. Ametller$^{a}$ and P. Talavera$^{bc}$}
\par\bigskip
$^a${\em
Departament de F\'\i sica i Enginyeria Nuclear,
Universitat Polit\`ecnica de Catalunya, \\
Jordi Girona 1Ð3, E-08034 Barcelona,  Spain}\\[.1cm]
$^b${\em
Departament de F{\'\i}sica i Enginyeria Nuclear,
Universitat Polit\`ecnica de Catalunya,\\
Comte Urgell 187, E-08036 Barcelona, Spain.}\\[.1cm]
$^c${\em
Institut de Ciencies del Cosmos,
Universitat de Barcelona,\\
Diagonal 647, E-08028 Barcelona, Spain.}\\[.1cm]
\vspace{5mm}
\end{center}

\section*{Abstract}

We show that a generalization of $su(2)$ Chiral Perturbation Theory,  including a perturbative singlet scalar field, converges faster towards the physical value of sensible low--energy observables.
The physical mass and width of the scalar particle are obtained through a simultaneous analysis of the pion radius and the 
$\gamma \gamma \to \pi^0\pi^0$ cross--section.  Both values are statistically consistent with the ones obtained by using Roy equations in $\pi-\pi$ scattering. In addition we find indications that the photon--photon--singlet coupling is quite small.

\vspace{3mm} \vfill{ \hrule width 5.cm \vskip 2.mm {\small
\noindent E-mail: lluis.ametller@upc.edu, ptal@mail.com }}

\newpage
\setcounter{page}{1}


\section{Motivation}

Scalar particles are neither too well known from experiment nor their properties are theoretically understood. Having the quantum numbers of the vacuum, the lightest scalar particle, the $\sigma$, couples strongly to pions and that makes it important for all the models involving spontaneous chiral symmetry breaking.

 Scalar extensions of Chiral Perturbation Theory ($\chi$PT) have been analyzed in the past by several authors. Recently, it has been proposed to include an isosinglet scalar as a dynamical degree of freedom to the $\chi$PT Lagrangian, \cite{Soto:2011ap} at the same footing as the lowest mass pseudoscalar Goldstone bosons.  With this, one is trying to obtain a better description of the low--energy processes among pions and, at the same time, to describe the nature of the $\sigma$. In fact it is commonly expected that the physics of the $\sigma$ would be governed by the dynamics of the Goldstone bosons, thus being the properties of the
interaction between two pions relevant \cite{Caprini:2005zr}.

In this note, we look for experiments involving pions and photons at low--energy which can be --at least in principle-- sensible to the dynamics of the scalar meson in order to obtain restrictions for the new couplings of the S$\chi$PT Lagrangian. In doing so, we restrict ourselves to the scenario where the free parameters of the model will be the $\sigma$ mass, $M_\sigma$, its width $\Gamma_\sigma$ and the $\sigma\pi\pi$ coupling constants. We focus on two observables: The vector form--factor of the charged pion and the $\gamma \gamma  \to \pi^0 \pi^0$ cross section. The first has been measured reasonably well in the space-like region and data for the second are rather old and also poor, but nevertheless it is a process of great interest thus can enlighten about the controversial $\sigma \gamma \gamma$ coupling.

A common feature for these two processes is that, in both, the number of additional constants wrt $\chi$PT is minimal. This is
given by the fact that in the aforementioned processes neither the mass nor the decay constant are renormalized at one loop. In addition, for the $\gamma \gamma \to \pi^0 \pi^0$ process not even the wave function renormalization is required. This essentially has as outcome that there is only one new coupling as relevant parameter. 

\section{Formalism }

Our starting point is the {S$\chi$}PT Lagrangian discussed in \cite{Soto:2011ap}, an extension of the lowest order {$\chi$}PT Lagrangian for Goldstone bosons with the inclusion of a scalar isosinglet. We will be concerned only with processes involving low--energy pions or photons as asymptotic states. In addition to this premise we should impose the scale hierarchy chain
\begin{equation}
p, M_\pi , M_{\sigma } \ll \Lambda_\chi\,.
\end{equation}
The S$\chi$PT Lagrangian involves pions and the $S_1$ scalar field --that we identify with the $\sigma$, transforming as a singlet under $SU(2)_L\times SU(2)_R$-- respects Chiral symmetry, $P$ and $C$ invariance and explicitly reads at lowest order
\begin{eqnarray}
\label{newsigmapion}
{\mathcal{L}}_{2}[0^{++}] = 
\left( \frac{F^2}{4} +F c_{1d} S_1+ c_{2d} S_1^2+\cdots \right) \langle u_\mu^\dagger  u^\mu \rangle+ \left( \frac{F^2}{4} + c_{2m} S_1^2+\cdots\right) \left( \langle \chi_+\rangle -\langle \chi^\dagger + \chi \rangle\right)\,.
\end{eqnarray}
Notice that by counting--power and gauge invariance a term involving the coupling $\gamma\gamma S_1$ is forbidden at this stage. As it stands (\ref{newsigmapion}) is a generalization of the Lagrangian corresponding to the singlet discussed in
\cite{Ecker:1988te} from where we borrow part of our notation in what follows.
The labels in the coupling constants $c$  indicate the number of scalar fields coupled to pions and the derivative- or massive-type of pion coupling. Ellipsis stand for higher order terms involving higher powers of the singlet field, which are scale suppressed. Here we take into account that $c_{1m}$ is zero, to enforce the scalar field to be a singlet under chiral symmetry and not mix with the vacuum.  As is customary the field $u$ parameterizes the pseudoscalar Goldstone bosons
\begin{equation}
u^2=U = e^{i\sqrt{2} \phi/F}\,,\quad
\phi= \begin{pmatrix}
\pi^0 & \sqrt{2} \pi^+\\
\sqrt{2} \pi^- & -\pi^0
\end{pmatrix}\,,
\end{equation}
and the $\chi$ field denotes the combination  
$\chi= 2 B_0\, (s+i p)$. $F$  is the pion decay constant in the chiral limit.  
In the rest we have made use of the following notation 
\begin{eqnarray}
&&u_\mu = i u^\dagger D_\mu U u^\dagger = -i u D_\mu U^\dagger u = u_\mu^\dagger\,,\nonumber\\
&&\chi_\pm = u^\dagger \chi u^\dagger \pm u \chi^\dagger u\,,\nonumber \\
&&f_\pm^{\mu\nu} = u F_L^{\mu\nu} u^\dagger \pm u^\dagger F_R^{\mu\nu} u\,.
\end{eqnarray}
The quantities $F_L^{\mu\nu}\,, F_R^{\mu\nu}$ are related with the field strength associated with the non--abelian external fields.

The $O(p^2)$ Lagrangian (\ref{newsigmapion}) will contribute to amplitudes at $O(p^4)$ through one--loop graphs, which in turn will give rise to ultraviolet divergences. In the case at hand the cancellation of such divergences proceeds only through a single counterterm, $\ell_6$\footnote{In order to avoid confusion with the low--energy constants  in $\chi$PT the S$\chi$PT ones are denoted by $\ell_i$ while
the former by $l_i$. }. In addition there can be a possible pure electromagnetic contribution in terms of a $S_1 \gamma\gamma$ coupling, $ c_{\gamma_1}$,
\begin{equation}
\label{sphotons}
{\mathcal{L}}_{4}[0^{++}] 
 =  \ell_6 {1\over 4} i \langle f_+^{\mu\nu} [u_\mu,u_\nu]\rangle\, + 
 \left( -{1\over 4} F_{\mu\nu} F^{\mu\nu}  - {\lambda \over 2} (\partial_\mu  A^\mu)^2 \right)(1 + c_{\gamma_1} S_1+\ldots )\,. 
\end{equation}
This last term has long been debated and is crucial to elucidate the composition of the scalar: non--strange $q\bar q$ state, $s\bar s$ state, 
tetra-quark state, $K\bar K$ molecule, glueball $\ldots$\,. However it is practically impossible to determine experimentally this coupling  at present and only
a combined study of several processes could in principle disentangle its value. For practical purpose we have considered it subleading in the counting--power, $\vert c_{\gamma_1}c_{1d}\vert  \le 1/(4 \pi F)^2$. Afterwards we will check the validity of this assumption through the consistency of the theoretical predictions versus the experimental results. We want to stress that this is the only point where  we add some extra assumption on top of just chiral symmetry constraints.

We have used dimensional regularization with $\omega\equiv  (d-4)/2$ in the $\overline{\text{MS}}$ scheme. In this regularization
the only low--energy constant we need is defined as:
\begin{equation}
\ell_6 = \ell_6^r + \gamma_6 \lambda\,,\quad 
\end{equation}
with 
$\lambda  = {\mu^{2\omega}\over 16 \pi^2}\left\{ {1\over 2 \omega}
-{1\over 2}(\log 4\pi + \Gamma^\prime(1) +1)\right\}
$. The $\ell_6^r$
is the coupling constant renormalized at the scale $\mu$
and the $\gamma_6$ factor is found via the Heat--Kernel expansion and is given by
\begin{equation}
\gamma_6= {1\over 3}(4 c_{1d}^2-1)\,.
\end{equation}
Finally the derivative of the $\Gamma$ function is the Euler constant, $\Gamma^\prime(1) = -\gamma$.

\subsection{The vector form--factor of the pion}

The vector form--factor of the pion is defined through the matrix element
$\langle\pi^i(p^\prime)\vert V_\mu^k\vert \pi^l(p)\rangle = i \epsilon^{ikl} (p^\prime_\mu+p_\mu) F_V(q^2)$. We have computed it at $O(p^4)$ in S$\chi$PT and have obtained the result 
\begin{eqnarray}
\label{fv}
F_V(t)=
1+\frac{t}{ {96 \pi^2 F^2}}  \left(\hbox{$\bar \ell_6$} -{1\over 3} \right)+ \frac{1}{6 F^2}(t-4 M_{\pi }^2)   
\hbox{$\bar{J}$}_{\pi\pi}(t) 
+{c_{1d}^2\over F^2} \frac{1}{(t-4 M_{\pi }^2)}   \left( P_V + U_V\right)\,.
\end{eqnarray}
The first three terms in (\ref{fv}) are independent of $c_{1d}$ and correspond to the well known $\chi$PT contribution \cite{Gasser:1983yg}, provided one identifies $\bar \ell_6$  with  $\bar l_6$. The rest is the contribution of the scalar singlet, and is split into two terms, a polynomial piece given by
\begin{eqnarray}
P_V &=&
- \frac{ \hbox{$\bar \ell_6$} t}{ {24 \pi^2 }} (t-4 M_{\pi }^2)
	-8 \left(M_{\sigma }^2-2 M_{\pi }^2\right)^2
\left[ \frac{ 4 M_\pi ^4+M_\sigma ^4-M_\pi ^2 (4 M_\sigma^2+t)}{ (M_\pi ^2-M_\sigma ^2)(4 M_\pi ^2-M_\sigma ^2 )}\right]  (\mu_\pi-\mu_\sigma)\\  
&+&\frac{\left[4 M_{\pi }^2 \left(14 M_{\pi }^2 t+72 M_{\pi }^4+t^2\right)+18 \left(12 M_{\pi }^2+t\right) M_{\sigma }^4- \left(68 M_{\pi }^2 t+432
   M_{\pi }^4+t^2\right) M_{\sigma }^2-36
   M_{\sigma }^6\right]}{ {72 \pi^2} \left(4 M_{\pi }^2-M_{\sigma }^2\right)}\,,\nonumber
\end{eqnarray}
and the dispersive part of the form--factor
\begin{eqnarray}
U_V &=&
4  \left(M_{\sigma }^2-2 M_{\pi }^2\right)^2 \left(4 M_{\pi}^2 - 2 M_{\sigma }^2-t\right) C_0\left(t,M_{\pi }^2,M_{\pi }^2,M_{\pi }^2,M_{\pi }^2,M_{\sigma }^2\right)\nonumber \\
  &-& \frac{2}{3} \left[ t^2 +64 M_{\pi}^4 +12 M_{\sigma }^4 -8 M_{\pi }^2
   \left(6 M_{\sigma }^2+t\right)  \right]  \hbox{$\bar{J}$}_{\pi \pi }(t)\\      
 &-&  4  \left(M_{\sigma }^2-2 M_{\pi }^2\right)^2 \left[  \frac{ 4 M_{\pi }^4+t M_{\sigma }^2 - M_{\pi }^2 \left(2 M_{\sigma}^2+3 t\right) }{M_{\pi }^2 
   \left(4 M_{\pi }^2-M_{\sigma }^2\right)} \right] \hbox{$\bar{J}$}_{\pi\sigma}(M_{\pi }^2)\,.
\end{eqnarray}
The $C_0\left(q^2,M_{\pi }^2,M_{\pi }^2,M_{\pi }^2,M_{\pi }^2,M_{\sigma }^2\right)$ function stands for the one--loop scalar three--point function \cite{Passarino:1978jh}, and $\bar{J}_{a b}( q^2)$ and $\mu_a$ are the one--loop scalar two--point and one--point function, subtracted at $q^2=0$, respectively \cite{Gasser:1983yg}.
The last term in (\ref{fv}) apparently contains a pole at $t=4 M_{\pi }^2$, but we have checked numerically that it is spurious. Moreover,
we have also checked numerically that, at zero momentum transfer, the
form--factor fulfills the expectations from the Ademollo--Gatto theorem \cite{AG}: $F_V (0^-) = 1$.
Notice that the above expression displays $t^2$ dependences that are customary of $O(p^6)$ in $\chi$PT.  That is the reason we
believe that S$\chi$PT can achieve a better convergence than $\chi$PT at moderate energies.

\subsection{The $\gamma\gamma\to \pi^0\pi^0$ amplitude}

It is well known that the lowest order contribution to this process in $\chi$PT is a pure $O(p^4)$ loop effect, which is finite by itself without any need of counterterms. This makes this process a gold plated test of $\chi$PT. However, when comparing the one loop prediction with existing experimental data, even near threshold, they differ significantly \cite{Donoghue:1988eea, Bijnens:1987dc}.  In order to improve the agreement, one is forced to work at two--loop order \cite{Bellucci:1994eb} or to rely on a dispersive treatment \cite{Morgan:1991zx}. We expect that the simple inclusion of the scalar particle would interpolate between both outcomes and ameliorate the situation at relatively higher energies, $\approx 0.6-0.7$ GeV. 

We have computed the $\gamma\gamma\to \pi^0\pi^0$ amplitude in S$\chi$PT assuming the mass scale hierarchy previously mentioned, where the direct $S_1\gamma\gamma$ coupling is negligible. This switches--off tree contributions and the process is driven entirely by loops making the comparison with $\chi$PT at the same footing.

The amplitude at $O(p^4)$ is purely S--wave and can be written as 
\begin{equation}
A(s)=  
 \frac{ 4 e^2}{F^2} {(s-M_{\pi }^2)\over s} \left[1  +F_\sigma(s)    \right]  \overline{\hbox{G}}(s) 
 \left( {1\over 2} s\, \epsilon_1\cdot \epsilon_2- q_1\cdot \epsilon_2\, q_2\cdot \epsilon_1\right),
 \label{ggpp}
 \end{equation}  
where $q_i$, $\epsilon_i$, are the photon momenta, polarizations respectively and  $s=(q_1+q_2)^2$.
We have collected the effects of the scalar singlet inside the factor
\begin{equation}
F_\sigma(s) =  -4 c_{1d}^2 \frac{(s-2M_\pi^2)^2}{(s-M_\pi^2)} \frac{1}{ s-s_0} \,, \quad s_0^{1/2}\equiv M_\sigma
\end{equation}
and finally the function $\overline{\hbox{G}}(s)$ is given in Eqs.(C1-C5) of Ref. \cite{Bellucci:1994eb}. Notice that we obtain a finite $O(p^4)$ amplitude.

For convenience when comparing with experimental results we will make use of the cross--section 
\begin{equation}
\label{cs}
{\sigma(\gamma\gamma\to\pi^0\pi^0)} = Z {\alpha^2 \pi \over F^4 }\frac{(s-M_\pi^2)^2}{ s}  \beta(s,M_\pi^2 ) \vert 1+F_\sigma (s)\vert^2 \vert  \overline{G}(s)\vert^2\,, \quad \beta(s,M_\pi^2 )= (1-4 M_\pi^2/ s)^{1/2}
\end{equation} 
where $Z$ is a factor that parameterizes the angular range of the experiment, $Z=\cos(\theta_{\hbox{max}})$. 

The most problematic feature involved in the previous expression (\ref{ggpp}) is that it does not comply with unitarity. In fact, there appears a pole at $s=s_0$.
In order to amend this drawback we 
regularize the real part by changing the above delta distribution by a Breit--Wigner,
\begin{equation}
\label{br}
s_0^{1/2}\to M_{\sigma }-i {\Gamma^\prime}\,.
\end{equation}
Even though the use of the Breit-Wigner distribution 
seems a bit controversial in our framework where $\Gamma^\prime \gg M_\sigma$ \cite{Kelkar:2010qn,Pennington:1999fa}.

There is substantial phenomenological evidence, \cite{Sannino:1995ik}, that $\Gamma^\prime$ {\sl can not} be interpreted 
as given directly in terms of the squared coupling constant. This would only be valid for a narrow resonance in a region where the background is negligible.
To circumvent this problem we consider $\Gamma^\prime$ as a phenomenological free parameter, at first instance unrelated to $c_{1d}$, checking afterwards the consistency of this picture.


\section{Numerical results}

In order to estimate the optimal values of the unknown parameters we have used a Monte--Carlo approach and fitted the available data on the space--like pion form--factor and on the $\gamma \gamma \to \pi^0\pi^0$ process to the corresponding theoretical expressions (\ref{fv}), (\ref{cs}).

Data for the $\gamma \gamma \to \pi^0\pi^0$ process are very scarce, relatively old and with very large uncertainties. We used a subset of
the Cristall Ball data \cite{Marsiske:1990hx} restricted to energies up to
$\approx 0.75~\text{GeV}$, where we expect our effective approach 
should still be valid.

For the pion form-factor we entirely rely on the space--like region data \cite{Dally:1982zk, Amendolia:1986wj}, but we have 
cross--checked that including the more fuzzy time--like data our findings are statistically consistent with the results we present below. 
When analyzed within $\chi$PT the two space--like data sets show a small inconsistency \cite{Bijnens:1998fm} that turns to be negligible for the sensitivity of our analysis.
The main reasons for focusing in the space--like region are: {\sl i)} the data set is large enough to evade some significant statistical fluke and {\sl ii)} the errors are rather small in comparison with available data in the time--like region.  As mentioned earlier the aim to include these data is to impose severe constraints on $c_{1d}$. In \cite{Soto:2011ap} this constant was obtained from the  decay width of the scalar by assuming that the latter is
obtained from Roy equations for the isoscalar S--wave $\pi \pi$ scattering amplitude near threshold  \cite{Caprini:2005zr}. 

The fitting strategy is as follows: we have randomly sampled with $2\times 10^6$ configurations the set of parameters
$\left\{ m_\sigma,\Gamma, \bar{\ell}_6,c_{1d}\right\}$  in the hypercube 
\begin{equation}
 0.3~\text{GeV} \le M_\sigma \le 0.7~\text{GeV}\,,\quad  0~\text{GeV} \le \Gamma^\prime \le 0.6~\text{GeV}\,,\quad 10 \le \bar{\ell}_6 \le 30\,,\quad 0.1 \le c_{1d} \le 0.5\,,\quad
\end{equation} 
with a priori flat distribution. The extremal values accommodate any reasonable outcome for those constants. 
The most favorable set of values is obtained by minimizing a $\chi^2$ distribution.
As numerical inputs we used the pion physical masses and  decay constant
\begin{equation}
M_{\pi^0}=134.9766\,\text{MeV}\,,\quad  M_{\pi^+}=139.57\,\text{MeV}\,,\quad F_\pi=93\,\text{MeV}\,. 
\end{equation}

\begin{figure}[h]
\centering
\includegraphics{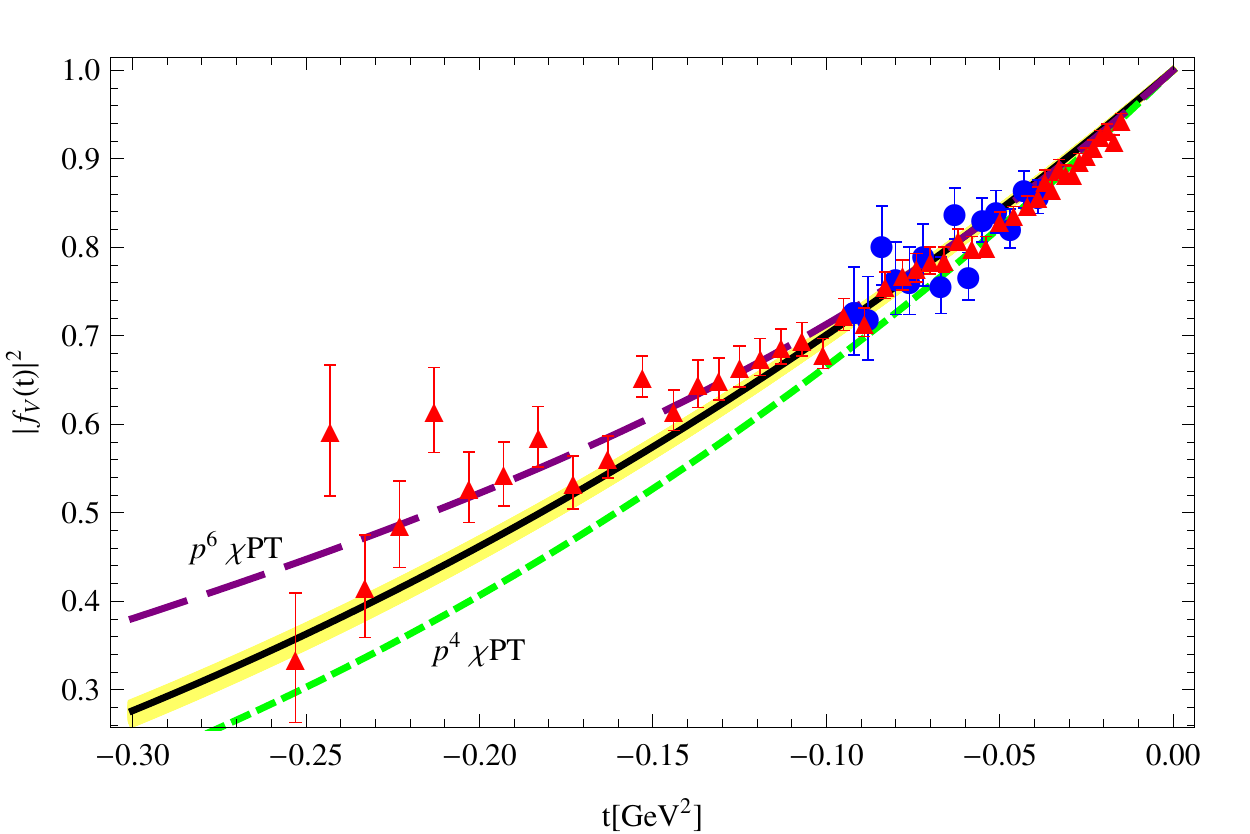}
\caption{Pion form--factor: Full line is obtained evaluating (\ref{fv}) at the central values of the parameters in (\ref{point}). 
For comparison we plot the $\chi$PT result at $O(p^4)$,  short--dashed line, 
with $\bar{l}_6= 16.5$ \cite{Gasser:1983yg}. If we decrease this value down to $\approx 15$ both curves, full and dashed, agree. 
The band covers the $1\sigma$ uncertainty around the values in (\ref{point}). The long--dashed line is the $\chi$PT result at $O(p^6)$ and is shown for completeness.}
\label{fig:fv}
\end{figure}

Before presenting the full analysis we perform individual fits for both processes with the result, 
\begin{enumerate}
\item $\pi\to \pi \gamma$
\begin{equation}
 \label{pointfv}
c_{1d}=0.10\,,\quad M_\sigma= 697~\text{MeV}\,,\quad \bar{\ell}_6=15.33\,,
 \end{equation}
\item $\gamma\gamma\to \pi^0\pi^0$
 \begin{equation}
 \label{pointgg}
c_{1d}=0.49\,,\quad M_\sigma= 413~\text{MeV}\,,\quad \Gamma^\prime= 399~\text{MeV}\,.
 \end{equation}
\end{enumerate}
The outcome is rather pedagogical: as the physics of the pion form--factor is already well understood in terms of vector saturation, the 
scalar contribution, if any, must be tiny. This is reflected in the small value of $c_{1d}$. Contrariwise, as the $\gamma\gamma \pi^0\pi^0$
cross--section is
very poorly understood in terms of pion rescattering effects this allows some room to incorporate the contribution of the scalar particle. 
We expect that the combined analysis maximizes the possible effect of the scalar particle in the  $\gamma\gamma \pi^0\pi^0$ reaction while we keep the common parameters under control due to the restrictions imposed by the pion form--factor.

The combined simultaneous fit is performed by minimizing an augmented $\chi^2$ distribution
 \begin{equation}
 \chi^2=\chi^2_{FV} + \chi^2_{\gamma \gamma\to \pi^0\pi^0}\,,
 \end{equation}
 where  both sets of experiments are weighted equally.
The landscape contains a single minimum for the $\chi^2$ function corresponding to 
 \begin{equation}
 \label{point}
c_{1d}=0.22_{-0.06}^{+0.13}\,,\quad \bar{\ell}_6=18.03_{-1.86}^{+10.39}\,,\quad 
M_\sigma= 497_{-64}^{+44}~\text{MeV}\,,\quad \Gamma^\prime= 233_{-117}^{+291}~\text{MeV}\,,\quad \chi^2_{\sl d.o.f}={162.7\over 65}\,.
 \end{equation}
The errors capture the deviation within 1$\sigma$ of the central result, i.e., we keep configuration points
 fulfilling 
$
 \chi^2 < \chi^2_{4,0.6827}
 $, where the upper bound corresponds to a probability of 68.27\% for the $4$ fitted parameters \cite{vuko}.
We have checked that, ballpark, any other point in a reasonable vicinity of (\ref{point}) leads to similar results.

\begin{figure}[h]
\centering
\includegraphics{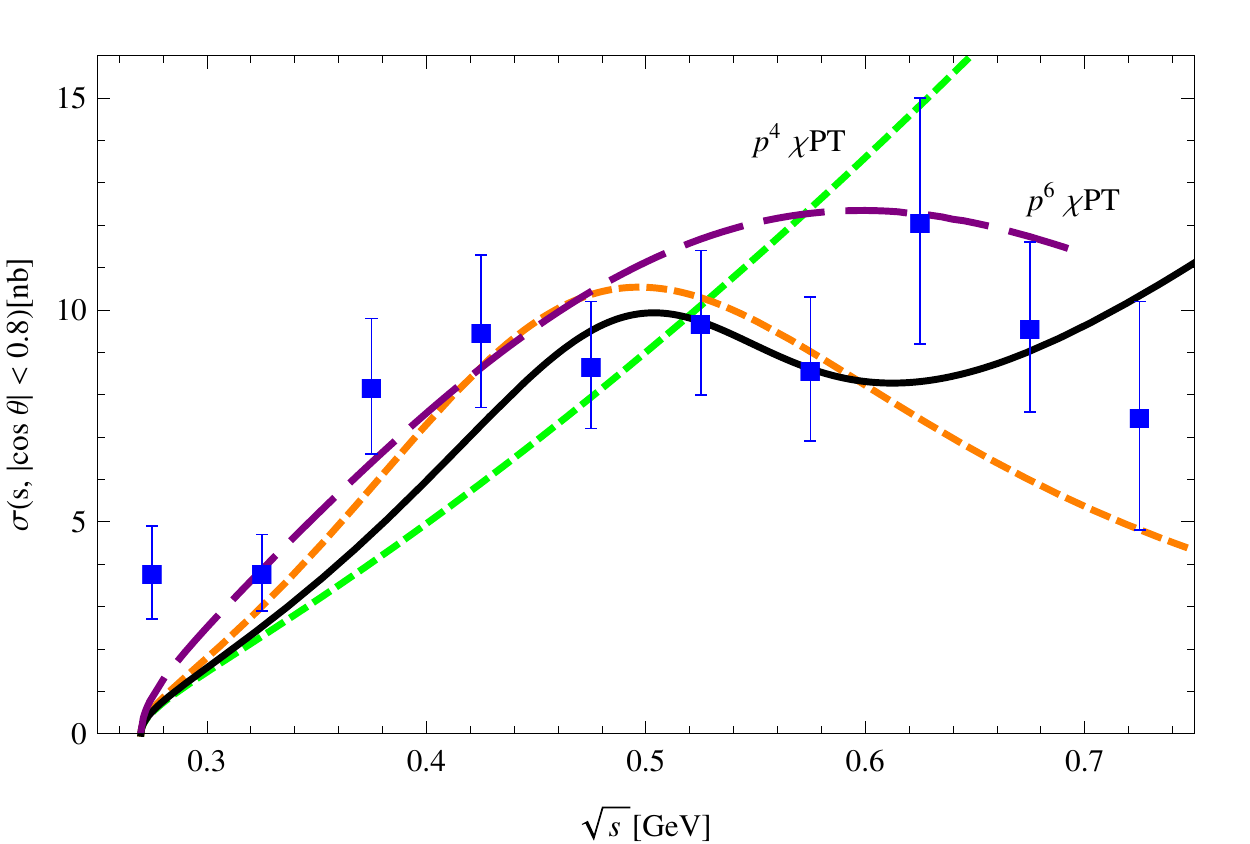}
\caption{Cross-section for $\gamma\gamma\to \pi^0\pi^0$ integrated over $\vert \cos{\theta^*}\vert < 0.8$ as a function of the $\pi\pi$ invariant mass $M_{\pi\pi}$: Full line is obtained evaluating (\ref{ggpp}) at (\ref{point}). The green short--dashed  line is the $O(p^4)$ $\chi$PT prediction \cite{Donoghue:1988eea} which does not depend on any low--energy constant. The long--dashed is the $O(p^6)$
$\chi$PT result in \cite{Gasser:2006qa}. The orange short--dashed curve shows the best  fit performed using the data of this channel alone (\ref{pointgg}).}
\label{fig:gg}
\end{figure}

In fig.(\ref{fig:fv})  and fig.(\ref{figggpp2}) we have plotted, full line, the solution corresponding to the central parameters (\ref{point}) together with the set of parameters that deviate from the former at most $1\sigma$, yellow band. For  comparison purposes we also depicted the $\chi$PT results in dashed lines, see fig.(\ref{fig:gg}). As one could anticipate the impact on the pion form--factor is almost imperceptible while the extra parameters wrt the $\chi$pt framework fairly accommodate the  $\gamma \gamma\to \pi^0\pi^0$  experimental data. It is worth noticing that the outcome of S$\chi$PT interpolates between $O(p^4)$ and $O(p^6)$ results of standard $\chi$PT.
 
\subsection{Comparison with earlier results}

Part of the outputs in (\ref{point}), $M_\sigma$ and $\Gamma^\prime$, can be compared to earlier results, see table 1. 
\begin{table}[h]
\label{sca}
\centering 
\begin{tabular}{c c c c c c c } 
&\cite{Caprini:2005zr}  & \cite{Aitala:2000xu} & \cite{Ablikim:2004qna} & \cite{GarciaMartin:2011jx}& \cite{Zhou:2004ms}&\cite{Bai:2004av} \\ [0.5ex] 
\hline 
\vspace{0.1cm}
$M_\sigma~\text{[MeV]}$&$441^{+16}_{-8}$ & $478\pm 29$ & $541\pm 29$ &  $457^{+14}_{-13}$&$470\pm 50 $&$434\pm 78$\\[1ex]
$\Gamma~\text{[MeV]}$&$544^{+18}_{-26}$ & $324\pm 22$& $504\pm 84$ & $558^{+22}_{-14}$ &$570\pm50 $&$404\pm 86$\\ [1ex] 
\hline 
\end{tabular}
\label{tableU} 
\caption{Comparison with some results in the literature.} 
\end{table}

\begin{figure}[htp]
\centering
\includegraphics{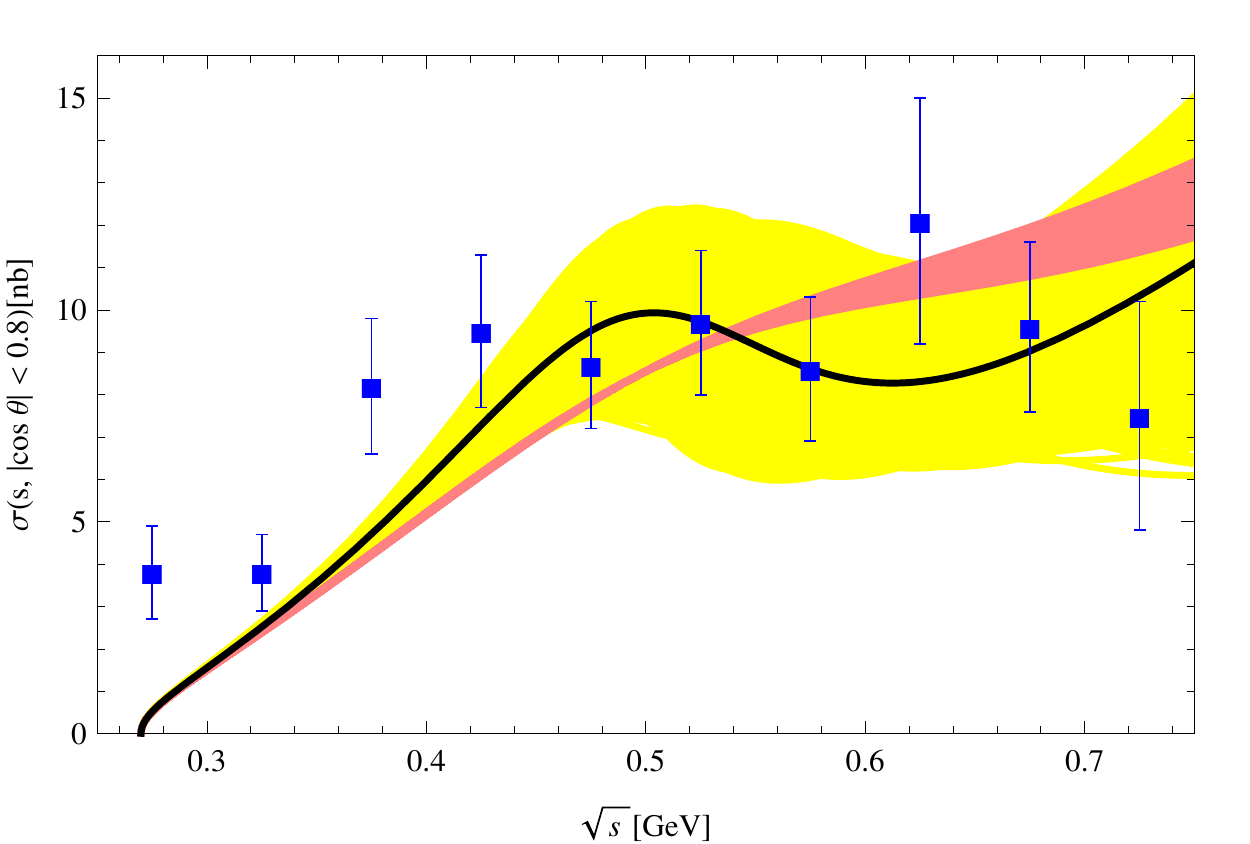}
\caption{Cross--section for $\gamma\gamma\to \pi^0\pi^0$ integrated over $\vert \cos{\theta^*}\vert < 0.8$ as a function of the $\pi\pi$ invariant mass $M_{\pi\pi}$: Full line is obtained evaluating (\ref{ggpp}) at the parameters (\ref{point}) and its $1\sigma$ deviation is covered
by the wider, yellow, band. The narrow pink band covers the $1\sigma$ deviation of (\ref{corr}).}
\label{figggpp2}
\end{figure}

\noindent While the agreement between masses is quite encouraging there is a mismatch between the central values for the widths that roughly amounts to a factor $2$,  with the exception of the data on $D\to \pi^+\pi^-\pi^+$ of E791 
\cite{Aitala:2000xu}.
In order to understand and quantify this disagreement, we fix the mass and the width of the $S_1$ to the central values given in \cite{Caprini:2005zr} and redo the analysis with the following outcome
\begin{equation}
\label{fix}
c_{1d}=0.26^{+0.005}_{-0.027}\,,\quad \bar{\ell}_6= 19.98^{+0.01}_{-1.32}\,,\quad \chi^2_{\sl d.o.f}={168.82\over 67}\,,
 \end{equation}
i.e. the central values of both $c_{1d}\,( \bar{\ell}_6)$ change by less than $20\%\, (10\%)$ with respect to the values in (\ref{point}). Furthermore both results are statistically equivalent, notice that the narrow pink band in fig.(\ref{figggpp2}) corresponding to the uncertainties 
on the central value of (\ref{fix}) is contained in the wider band corresponding to the $1\sigma$ values of (\ref{point}) while the deviation in the pion form--factor is still within the $1\sigma$ band in fig.(\ref{fig:fv}). In view of the previous result one can conclude that, with the present data on $\gamma\gamma\pi^0\pi^0$,
our results for the mass and width of the salar field (\ref{point}) are compatible with that in \cite{Caprini:2005zr}. 

The value of the constant $c_{1d}$ must be compared with 
that obtained in \cite{Soto:2011ap}, $c_{1d}=0.67$, where lattice data were used and, more important, the coupling was essentially deduced from the scalar decay width. In the present analysis this constant turns out to be a factor $3$ smaller.  As mentioned above  the assumption that the physical width and coupling constant of the vertex $\sigma\pi\pi$ are related by a simple dispersion relation is probably too naive.  Even-though one should bear in mind that, as is evident from the form of (\ref{fv}), there must be a strong linear correlation between the pair
$\{c_{1d}^2,\bar{\ell}_6\}$.

To evaluate this statement quantitatively we have evaluated the correlation matrix between the different pairs of variables
\begin{equation}
\label{corr}
  \bordermatrix{&c_{1d}^2&\bar{\ell}_6&M_\sigma^2&\Gamma^\prime \cr
c_{1d}^2&    1& 0.97 &-0.26  &0.72 \cr
\bar{\ell}_6&   0.97 & 1 & -0.24 & 0.68  \cr
M_\sigma^2&    -0.26 & -0.24& 1 &-0.3    \cr
\Gamma^\prime&    0.72 & 0.68 & -0.3 &  1 }\,.
\end{equation}
As a consequence one can increase the value of $c_{1d}$ by increasing  $\bar{\ell}_6$ but, nevertheless, the value obtained in 
\cite{Soto:2011ap} for $c_{1d}$ is so big that its corresponding $\bar{\ell}_6$ implied for (\ref{corr}) is presumably ruled out by some direct measurement of the pion radii.
  
Finally, the value of $\bar{\ell}_6$ in (\ref{point}) 
can be crosschecked with the two--loop one obtained in \cite{Bijnens:1998fm}  $\bar{l}_6= 16\pm 0.5 \pm 0.7$.

\section{A sample of applications}
Once our main results are obtained, we discuss their implications in a sample of applications. From one side, the dynamical scalar will contribute to some pion properties, such as the neutral pion scattering lengths and pion polarizabilities. On the other side, the $\sigma$ itself gets effective radiative couplings at one loop that translate in a non vanishing $\sigma\to\gamma\gamma$ decay. 

\subsection{$\pi\pi$ scattering lengths}
The picture for the $\pi\pi$ scattering lengths which emerges out of our low--energy Lagrangian results in a new contribution to the  Current Algebra (CA), coming from the scalar in the $s$--channel at tree--level, and proportional to $c_{1d}$. Provided this is tiny we expect no
huge departures from CA. It explicitely reads
\begin{eqnarray}
a^0_0 &=&\frac{M_{\pi }^2}{32 \pi  F^2} \left[7+16  {c_{1d}}^2
   \left(3 \frac{M_{\pi }^2}{M_{\sigma }^2 -4 M_{\pi }^2}+2\frac{M_{\pi }^2}{ M_{\sigma
   }^2}\right)\right]\,,\\
a^2_0 &=& \frac{M_{\pi }^2}{16
   \pi  F^2} \left(-1+16 {c_{1d}}^2 \frac{ M_{\pi }^2}{ M_{\sigma }^2}\right)\,,
\end{eqnarray}
whose values are collected in the table 2.
\begin{table}[h]
\label{Csca}
\centering 
\begin{tabular}{c c c c c } 
&CA  & S$\chi$PT & $O(p^4) \chi$PT& \text{Ex.(stat)(syst)} \\ [0.5ex] 
\hline 
\vspace{0.1cm}
$a_0^0$&0.158 & $0.168^{+0.015}_{-0.007}$ & 0.2 &  0.2210(47)(40)\\[1ex]
$a_0^2$&-0.045 & $-0.042^{+0.005}_{-0.001}$& - 0.042 & -0.0429(44)(28) \\ [1ex] 
\hline 
\end{tabular}
\caption{Comparison for the scattering--lengths.} 
\end{table}

As one can appreciate the S$\chi$PT results nicely interpolate once more between two consecutive $\chi$PT order results. 
We have checked that all values within the $1\sigma$ deviation for the scattering lengths in the above table are inside the universal band as defined in \cite{Ananthanarayan:2000ht}.

\subsection{Neutral pion polarizabilities}
To obtain the pion polarizabilities we consider the crossed channel $\gamma \pi^0\to \gamma \pi^0$ at threshold. In our case, see (\ref{ggpp}),  the electric $(\overline{\alpha}_{\pi^0})$ and magnetic $(\overline{\beta}_{\pi^0})$ polarizabilities are identical to each other. 
Introducing a $4\pi$ factor to conform the experimental data we obtain
\begin{equation}
\label{polas}
\overline{\alpha}_{\pi^0}- \overline{\beta}_{\pi^0}= -{\alpha\over 3  \pi^2 F^2  M_{\pi^0}} \lvert {1\over 16} - c_{1d}^2  {M_{\pi^0}^2\over M_\sigma^2-i M_\sigma\Gamma^\prime}\rvert = \left(-1.01 +  0.03 \pm 0.001\right) \times 10^{-4} {\text fm}^3\,,
\end{equation}
where the first quantity is the $\chi$PT contribution, the second is the scalar contribution and the errors are the maximum and minimum deviation inside the 1$\sigma$ values of (\ref{point}). 
The previous result must be compared with the experimental one $\left(\alpha_{\pi^0}- \beta_{\pi^0}\right)^{\text{exp}}=-1.1\pm1.7$ \cite{Kaloshin:1993wj}. 
Moreover the correction due to the scalar particle in (\ref{polas}) is roughly a factor $20$ smaller than the two--loop 
expression \cite{Bellucci:1994eb} meaning that the polarizabilities measurements by themselves neither would verify the existence of the scalar particle nor determine its characteristics.

\subsection{$S_1$ radiative width}

As far as the scalar particle is concerned, although we have explicitly supressed its direct coupling to photons at leading order, our scheme allows a dynamically generated $\gamma\gamma S_1$ interaction, via pion loops at $O(p^4)$. In this context one obtains
 \begin{equation} 
 \label{swidh}
\Gamma_{S_1\to\gamma\gamma} = 16 \pi \alpha^2 {c_{1d}^2 \over F^2}\frac{(M_\sigma^2-2M_\pi^2)^2}{M_\sigma}
\lvert{\overline G}(M_\sigma^2)\rvert^2= 0.11~\text{KeV}\,.
\end{equation} 
This result lies somewhat below the lower edge of the range $[0.22, 4.4]\,$ KeV that is available in the literature, see Table 1 in \cite{Pennington:2007yt}. Even-though
we expect at this stage that a 
 direct $S_1\gamma\gamma$ coupling terms coming from (\ref{sphotons})  would give contributions numerically of the same order as those in (\ref{swidh}). In view of the previous numerical result it seems hard to reconciliate the picture of the singlet with a simple
 $u\bar{u}, d\bar{d}$ composition as found in \cite{Pennington:2006dg}. 

\subsection{Hadronic contribution to Muon $(g-2)$ and to $\alpha(M_Z^2)$}

We reevaluate the hadronic contribution to the running of the QED fine structure constant $\alpha(s)$ at $s=M_Z^2$ and the contribution from
hadronic vacuum polarization. Using analyticity and unitary of the vacuum polarization correlator both contributions can be calculated via dispersion integrals 
\begin{eqnarray}
 & &a_\mu^{\text had}= {\alpha_{QED} ^2\over 3\pi^2}\int_{4 M_\pi^2}^\infty {ds\over s} R(s) K(s)\quad\quad\, \cite{CG}\,, \nonumber\\
 & &\Delta\alpha(M_Z^2)=-{\alpha_{QED}\over 3 \pi} M_Z^2  \mathbf{Re} \int _{4 M_\pi^2}^\infty ds {R(s)\over s(s-M_Z^2-i \epsilon)}\quad\quad\,
\cite{GD}\,,
\end{eqnarray}
where $K(s)$ is the QED kernel \cite{Eidelman:1995ny}.
In turn 
both magnitudes are related via dispersion relation to the hadronic production rate in $e^+ e^-$ annihilation. Assuming that the main contribution of the latter at low--energies is given entirely by the pion contribution, one obtains
\begin{equation}
R(s)={\sigma(e^+e^-\to {\text {hadrons}})\over \sigma(e^+e^-\to {\mu^+\mu^-})}  \approx {1\over 4}\beta(s,M_\pi^2)^3 \vert F_V(s)\vert^2\,.
\end{equation}
Obviously the main contribution to $F_V$ is dominated by the $\rho(770)$ but at energies below $500$~MeV there is a considerable fraction coming from the scalar resonance that can compete with the $\rho(770)$ tail. Inserting (\ref{fv}) in $F_V$ above 
the contributions to both
quantities as a function of the cutoff $\Lambda$ are given in table 3. Once more the results including the singlet effects interpolate between
two consecutive chiral orders.
\begin{table}[h] \begin{center}
\begin{tabular}{cccccccc}
\multicolumn{1}{c}{} & \multicolumn{3}{c}{$10^{-10} \times a^{\text {had}}_\mu$ }
&\multicolumn{1}{c}{} &\multicolumn{3}{c}{$10^{-4} \times \Delta\alpha(M_Z^2)$} \\
\multicolumn{1}{c}{$\Lambda$({\text GeV})} & \multicolumn{1}{c}{$p^4\, \chi PT$ } & \multicolumn{1}{c}{$S\chi PT$}& \multicolumn{1}{c}{$p^6\, \chi PT$ \cite{Bijnens:1998fm}}&
\multicolumn{1}{c}{} &  \multicolumn{1}{c}{$p^4\, \chi PT$}
 &\multicolumn{1}{c}{$S\chi PT$} & \multicolumn{1}{c}{$p^6\, \chi PT$ \cite{Bijnens:1998fm}}\\
 \hline
\multicolumn{1}{c}{$0.32$} & \multicolumn{1}{c}{$2.12$ } & \multicolumn{1}{c}{$2.26$}&\multicolumn{1}{c}{$2.38$}&\multicolumn{1}{c}{} & \multicolumn{1}{c}{$0.035$}
 &\multicolumn{1}{c}{$0.037$}&\multicolumn{1}{c}{$0.039$} \\
 \multicolumn{1}{c}{$0.35$} & \multicolumn{1}{c}{$6.48$ } & \multicolumn{1}{c}{$6.92$}& \multicolumn{1}{c}{$7.4$} &\multicolumn{1}{c}{} & \multicolumn{1}{c}{$0.117$}
 &\multicolumn{1}{c}{$0.125$}&\multicolumn{1}{c}{$0.13$} \\
 \multicolumn{1}{c}{$0.40$} & \multicolumn{1}{c}{$16.76$ } & \multicolumn{1}{c}{$18.11$}& \multicolumn{1}{c}{$20.0$} &\multicolumn{1}{c}{} & \multicolumn{1}{c}{$0.350$}
 &\multicolumn{1}{c}{$0.379$}&\multicolumn{1}{c}{$0.42$} \\
 \multicolumn{1}{c}{$0.45$} & \multicolumn{1}{c}{$28.62$ } & \multicolumn{1}{c}{$31.25$}& \multicolumn{1}{c}{$35.7$}&\multicolumn{1}{c}{} & \multicolumn{1}{c}{$0.679$}
 &\multicolumn{1}{c}{$0.744$}&\multicolumn{1}{c}{$0.86$} \\
  \multicolumn{1}{c}{$0.50$} & \multicolumn{1}{c}{$40.72$ } & \multicolumn{1}{c}{$44.92$}& \multicolumn{1}{c}{$53.6$}&\multicolumn{1}{c}{} & \multicolumn{1}{c}{$1.084$}
 &\multicolumn{1}{c}{$1.200$}&\multicolumn{1}{c}{$1.45$} \\
\hline
\end{tabular}
\caption{$a^{\text {had}}_\mu$ and  $\Delta\alpha(M_Z^2)$ as a function of the cutt-off $\Lambda$. } 
\end{center}
\end{table}
Those results must be compared with the experimental results $a^{\text {had}}_\mu = (695.1\pm7.5)\times 10^{-10}\,,\Delta\alpha(M_Z^2) = (277.8\pm 2.6)\times 10^{-4}$ \cite{Davier:1997kw}. Notice that at $\Lambda = 0.5$ GeV the difference between the
$O(p^4)\, \chi PT$ and $S\chi PT$ in both quantities roughly amounts to half the experimental error.

\subsection{Pion radii}

We can now expand the form factor (\ref{fv}) for $t \ll 4 M_\pi^2$ and obtain the expression 
\begin{equation}
F_V = 1 +{1\over 6} \langle r^2\rangle_{V}^\pi t +\ldots
\end{equation}
where the pion charge radius is given by the linear terms as
\begin{eqnarray}
&&\langle r^2\rangle_{V}^\pi= {1\over 16\pi^2 F^2} (\bar{\ell}_6-1)-3 {c_{1d}^2\over F^2}\left[ {1\over 48 \pi^2 }{M_{\sigma }^2 \over M_{\pi }^4} (M_{\sigma }^2- 4 M_{\pi }^2) + {1\over 12\pi^2 } \bar{\ell}_6
\right. \nonumber \\ &&
\left. + {\left(M_{\sigma }^2-2 M_{\pi }^2\right)^2\over M_{\pi }^4} \left\{  
{M_{\sigma }^2 \over  (M_{\pi }^2-M_{\sigma }^2) } (\mu_\pi-\mu_\sigma) 
+\hbox{$\bar{J}$}_{\pi\sigma}(M_{\pi }^2) - M_{\sigma }^2 C_0\left(t,M_{\pi }^2,M_{\pi }^2,M_{\pi }^2,M_{\pi }^2,M_{\sigma }^2\right) 
\right. \right. \nonumber \\ &&
\left. \left. 
- 4 M_{\pi }^4 \left(M_{\sigma }^2-2 M_{\pi }^2\right) \partial_t C_0\left(t,M_{\pi }^2,M_{\pi }^2,M_{\pi }^2,M_{\pi }^2,M_{\sigma }^2\right)
{ \over }\right\}\rvert_{t\to 0^+} \right]\,.
\end{eqnarray}
We refrain of evaluating the previous expression numerically because it contains instabilities in the kinetic range it is defined.

\section{Final remarks}
We have considered $\chi$PT enlarged with the inclusion of a scalar particle as a dynamical degree of freedom. By fitting experimental data on the vector form--factor of the pion and on the $\gamma\gamma\to\pi^0\pi^0$ cross--section, we have extracted the favoured values for the $S_1\pi\pi$ coupling constant, the mass and width of the scalar  particle and a low--energy constant $\bar{\ell}_6$ finding that for the mass and decay width the results are statistically equivalent to those extracted with high-energy data.
We have analyzed and computed the effects of this particle on a wide set of data and have found that they are somewhere in between the predictions of two consecutive orders in $\chi$PT, what makes the framework a useful extension in parameter range of the predictions of $\chi$PT.

\subsection{Acknowledgments}
We are grateful to J.~Bijnens, J.~Gasser and M.~Ivanov for providing the data for the two--loop curves in our figures and to Ll.~Garrido
for discussion about some statistics issues.

PT gratefully acknowledges support from FPA2010-20807, 2009SGR502 and Consolider grant
CSD2007-00042 (CPAN).

\bibliographystyle{ssg}
\bibliography{hair}

\end{document}